\documentclass[preprint]{epl}

\usepackage{amsmath} 

\def\bk#1{\langle#1\rangle}
\def\dbk#1{\langle\!\langle#1\rangle\!\rangle}
\def\Bigdbk#1{\Big\langle\!\!\Big\langle#1\Big\rangle\!\!\Big\rangle}
\def\sgn{{\rm sgn}}
\def\del{\nabla}
\def\pd{{\phantom{\dagger}}}
\def\sf{\text{B}}
\def\so{\text{so}} 
\def\myRe{\text{Re}}
\def\myIm{\text{Im}}

\title{Enhancing superconductivity: Magnetic impurities 
and their quenching by magnetic fields}

\shorttitle{Enhancing superconductivity}

\author{T.-C. Wei
\and 
D. Pekker 
\and
A. Rogachev 
\and 
A. Bezryadin 
\and
P. M. Goldbart} 

\shortauthor{T.-C. Wei et al.}

\institute{Department of Physics, University of Illinois at
Urbana-Champaign, \\ 1110 West Green Street, Urbana,  Illinois 61801-3080,
U.S.A.}

\pacs{74.25.Sv}{Critical current} 
\pacs{74.62.-c}{Transition temperature variations}
\pacs{74.78.-w}{Superconducting films and low-dimensional structures}

\begin{document}

\maketitle

\begin{abstract}
Magnetic fields and magnetic impurities are each known to suppress
superconductivity.  However, as the field quenches (i.e.~polarizes)
the impurities,
rich consequences, including field-enhanced superconductivity,
can emerge when both effects are present.
For the case of  superconducting wires and thin films, 
this field-spin interplay is investigated via the
Eilenberger-Usadel scheme.
Non-monotonic dependence of the critical current on the
field (and therefore field-enhanced superconductivity) 
is found to be possible, even in parameter regimes
in which the critical temperature decreases monotonically with increasing
field.  The present work complements that of Kharitonov and
Feigel'man, which predicts  non-monotonic behavior of the
critical temperature.
\end{abstract}

\section{Introduction}
In their classic work, Abrikosov and Gor'kov~\cite{AbrikosovGorkov60}
predicted that unpolarized, uncorrelated magnetic impurities suppress
of superconductivity, due to the de-pairing effects associated with the
spin-exchange scattering of electrons by magnetic impurities.  Among
their results is the reduction, with increasing magnetic impurity
concentration, of the superconducting critical temperature $T_{\rm
c}$, along with the possibility of \lq\lq gapless\rq\rq\
superconductivity in an intermediate regime of magnetic-impurity
concentrations. The latter regime is realized when the concentration of the
impurities is large enough to eliminate the gap but not large enough
to destroy superconductivity altogether. Not long after the work of
Abrikosov and Gor'kov, it was recognized that other de-pairing
mechanisms, such as those involving the coupling of the orbital and
spin degrees of freedom of the electrons to a magnetic field, can lead
to equivalent suppressions of superconductivity, including gapless
regimes~\cite{Maki64,deGennesTinkham64,MakiFulde65,FuldeMaki66}.

Conventional wisdom holds that magnetic fields and magnetic moments
each tend to suppress 
superconductivity (see, e.g., Ref.~\cite{deGennesTinkham}).
Therefore, it seems natural to suspect that any increase in a magnetic
field, applied to a superconductor containing magnetic impurities,
would lead to additional suppression of the superconductivity.
However, very recently, Kharitonov and
Feigel'man~\cite{KharitonovFeigelman05} have predicted the
existence of a regime in which, by contrast, an increase in the
magnetic field applied to a superconductor containing magnetic
impurities leads to a  critical temperature that first increases
with magnetic field, but eventually behaves more conventionally,
decreasing with the magnetic field and ultimately vanishing at a
critical value of the field.  Even more strikingly, they have
predicted that, over a certain range of concentrations of magnetic
impurities, a magnetic field can actually induce superconductivity out
of the normal state.

The Kharitonov-Feigel'man treatment focuses on determining the
critical temperature by determining the linear instability of the
normal state.  The purpose of the present Letter is to address
properties of the superconducting state itself, most notably the
critical current and its dependence on temperature and the externally
applied magnetic field.  The approach that we shall take is to derive
the (transport-like) Eilenberger-Usadel
equations~\cite{Eilenberger66,Usadel70}, by starting from the Gor'kov
equations.  We account for the following effects: potential and
spin-orbit scattering of electrons from non-magnetic impurities, and
spin-exchange scattering from magnetic impurities, along with orbital
and Zeeman effects of the magnetic field.  In addition to obtaining
the critical current, we shall recover the Kharitonov-Feigel'man
prediction for the critical temperature, as well as the dependence of
the order parameter on temperature and applied field.  In particular,
we shall show that not only are there reasonable parameter regimes in
which both the critical current and the transition temperature vary
non-monotonically with increasing magnetic field, but also there are
reasonable parameter regimes in which only the low-temperature
critical current is non-monotonic even though the critical temperature
behaves monotonically with field.  The present theory can be used to
explain certain recent experiments on superconducting
wires~\cite{RogachevEtAl}.

Before describing the technical development, we pause to give a
physical picture of the relevant de-pairing mechanisms. First,
consider the effects of magnetic impurities. These cause spin-exchange
scattering of the electrons (including both spin-flip and
non-spin-flip terms, relative to a given spin quantization axis), and
therefore lead to the breaking of Cooper
pairs~\cite{AbrikosovGorkov60}. Now consider the effects of magnetic
fields.  The vector potential (due to the applied field) scrambles the
relative phases of the partners of a Cooper pair, as they move
diffusively in the presence of impurity scattering (viz.~the orbital
effect), which suppresses
superconductivity~\cite{Maki64,deGennesTinkham64}.  On the other hand,
the field polarizes the magnetic impurity spins, which
decreases the rate of exchange scattering (because the spin-flip term
is suppressed more strongly than the non-spin-flip term is enhanced),
thus diminishing this contribution to
de-pairing~\cite{KharitonovFeigelman05}.  In addition, the Zeeman
effect associated with the effective field (coming from the applied
field and the impurity spins) splits the energy of the up and down
spins in the Cooper pair,  thus tending to suppress
superconductivity~\cite{deGennesTinkham}. We note that strong spin-orbit
scattering tends to weaken the de-pairing caused by the Zeeman
effect~\cite{FuldeMaki66}.  Thus we see that the magnetic field
produces competing tendencies: it causes de-pairing via the orbital
and Zeeman effects, but it mollifies the de-pairing caused by magnetic
impurities.  This competition can manifest itself through the
non-monotonic behavior of observables such as the critical temperature
and critical current. In order for the manifestation to be observable,
the magnetic field needs to be present throughout the samples, the
scenario being readily accessible in wires and thin films.

\section{The model}
We take for the impurity-free part of the Hamiltonian the BCS mean-field form~\cite{BCS,deGennesTinkham}:
\begin{equation}
H_0=-\int dr\,\frac{1}{2m}\psi^\dagger_\alpha \Big(\del-\frac{ie}{
c}A\Big)^2\psi^\pd_\alpha + \frac{V_0}{2}\int dr \,
\left(\bk{\psi^\dagger_\alpha\psi^\dagger_\beta}\psi_\beta\psi_\alpha+
\psi^\dagger_\alpha\psi^\dagger_\beta\bk{\psi_\beta\psi_\alpha}\right)-
\mu\int dr\, \psi^\dagger_\alpha\psi_\alpha,
\end{equation}
where \(\psi^\dagger_\alpha(r)\) creates an electron having mass
\(m\),charge $e$, position \(r\) and spin projection \(\alpha\), $A$
is the vector potential, $c$ is the speed of light, \(\mu\)
is the chemical potential, and \(V_0\) is the pairing interaction.
Throughout this Letter we shall put $\hbar=1$ and $k_B=1$. Assuming
the superconducting pairing is spin-singlet, we may introduce the
complex order parameter $\Delta$, via
\begin{align}
 -V_0\bk{\psi_\alpha\psi_\beta}
&=i\sigma^y_{\alpha\beta}\Delta,&
 V_0\bk{\psi^\pd_\alpha\psi^\pd_\beta}
&=i\sigma^y_{\alpha\beta}\Delta^*,
\end{align}
where \(\sigma^{x,y,z}_{\alpha \beta}\) are the Pauli matrices.  We
assume that the electrons undergo potential and spin-exchange
scattering from the magnetic impurities located at a set of random
positions
$\{x_i\}$, in addition to undergoing spin-orbit
scattering from an independent set of impurities or defects located
at an independent set of random positions $\{y_j\}$, as well as being
Zeeman coupled to the applied field:
\begin{subequations}
\begin{equation}
H_{\rm int}=\int dr\, \psi^\dagger_\alpha V^\pd_{\alpha\beta} \psi^\pd_\beta,
\end{equation}
with $V_{\alpha\beta}$ being given by
\begin{equation}
V_{\alpha\beta}=\sum\nolimits_{i}\big\{u_1(r\!-\!x_i)\delta_{\alpha\beta}+
u_2(r\!-\!x_i)\vec{S}_i\cdot\vec{\sigma}_{\alpha\beta}\big\}+
{\sum}_j\big\{\vec{\del} v_{so}(r-y_j)\cdot \big(\vec{\sigma}_{\alpha\beta}\times \vec{p}\big)\big\}+
\mu_B B\,\sigma^z_{\alpha\beta},
\label{Vab}
\end{equation}
\end{subequations}
where $\vec{S}_i$ is the spin of the $i$-th magnetic impurity and
where, for simplicity, we have attributed the potential scattering
solely to the magnetic impurities.  We could have included potential
scattering from the spin-orbit scattering centers, as well as
potential scattering from a third, independent set of impurities.
However, to do so would not change our conclusions, save for the
simple rescaling of the mean-free time.  We note that cross terms,
i.e. those involving distinct interactions, can be ignored when
evaluating self-energy~\cite{FuldeMaki66,KharitonovFeigelman05}.
Furthermore, we shall assume that the Kondo temperature is much
lower than the temperature we are interested in.

The impurity spins interact with the applied magnetic field through
their own Zeeman term:
\begin{equation}
H_{\rm Z}=-\omega_s S^z
\end{equation}
where $\omega_s\equiv g_s\mu_B B$, and $g_s$ is the impurity-spin
$g$-factor. Thus, the impurity spins are not treated as static but
rather have their own dynamics, induced by the applied magnetic field.
We shall approximate the dynamics of the impurity spins as being
governed solely by the applied field, ignoring any influence on them
of the electrons.  Then, as the impurity spins are in thermal
equilibrium, we may take the Matsubara correlators for a single spin
to be
\begin{subequations}
\begin{eqnarray}
 \bk{ T_\tau S^+(\tau_1) S^-(\tau_2)}&=&
T{\sum}_{\omega'} D^{+-}_{\omega'} e^{-i\omega'(\tau_1-\tau_2)},\\
 \bk{ T_\tau S^-(\tau_1) S^+(\tau_2)}&=&
T{\sum}_{\omega'} D^{-+}_{\omega'} e^{-i\omega'(\tau_1-\tau_2)},\\
\bk{ T_\tau S^z(\tau_1) S^z(\tau_2)}&=&
 d^{z}=\overline{(S^z)^2},
\end{eqnarray}
\end{subequations}
where $\omega'$ ($\equiv 2\pi n T$) is a bosonic Matsubara frequency,
$\overline{\cdots}$\, denotes a thermal average, and
\begin{align}
D^{+-}_{\omega'}&\equiv {2\overline{S^z}}/({-i\omega'+\omega_s}),& 
D^{-+}_{\omega'}&\equiv {2\overline{S^z}}/({+i\omega'+\omega_s}).
\end{align}
We shall ignore correlations between distinct
impurity spins, as their effects are of the second order in the impurity
concentration.  

To facilitate the forthcoming analysis, we define the Nambu-Gor'kov four-component spinor (see, e.g., Refs.~\cite{FuldeMaki66,AmbegaokarGriffin65}) via
\begin{equation} 
    \Psi^\dagger(x)\equiv
    \Big(\psi^\dagger_\uparrow(r,\tau),
      \psi^\dagger_\downarrow(r,\tau),
      \psi_\uparrow(r,\tau),
      \psi_\downarrow(r,\tau)\Big).
\end{equation}
Then, the electron-sector Green functions are defined in the standard way via
\begin{equation}
{\cal G}_{ij}(1:2)\equiv - \langle T_\tau \Psi_i(1)\Psi^\dagger_j(2) \rangle
\equiv\begin{pmatrix}
   \hat{G}(1:2) & \hat{F}(1:2) \cr
   \hat{F}^\dagger(1:2) & \hat{G}^\dagger(1:2)
 \end{pmatrix},
\end{equation}
where $\hat{G}$, $\hat{G}^\dagger$, $\hat{F}$, and $\hat{F}^\dagger$
are each two-by-two matrices (as indicated by the $\,\hat{}\,$ symbol), being
the normal and anomalous Green functions, respectively. As the pairing
is assumed to be singlet, $\hat{F}$ is off-diagonal whereas $\hat{G}$
is diagonal.

\section{Eilenberger-Usadel equations}
The critical temperature and critical current are two of the most
readily observable quantities.  As they can be readily obtained
from the Eilenberger and Usadel equations, we shall focus on these
formalisms.  A detailed derivation will be presented elsewhere. The
procedure is first to derive Eilenberger
equations~\cite{Eilenberger66}, and then, assuming the dirty limit, to
obtain the Usadel equations. The self-consistency equation between the
anomalous Green function and the order parameter naturally leads, in
the small order-parameter limit, to an equation determining the
critical temperature. Moreover, solving the resulting transport-like
equations, together with the self-consistency equation, gives the
transport current, and this, when maximized over superfluid velocity,
yields the critical current.

To implement this procedure, one first derives the equations of motion
for ${\cal G}$ (viz.~the Gor'kov equations).  By suitably subtracting
these equations from one another one arrives at a form amenable to a
semiclassical analysis, for which the rapidly and slowly varying
parts in the Green function (corresponding to the dependence on the
relative and center-of-mass coordinates of a Cooper pair,
respectively) can be separated. Next, one treats the interaction
Hamiltonian as an insertion in the self-energy, which leads to a new set of
semi-classical Gor'kov equations.  These equations are still too
complicated to use effectively, but they can be simplified to the so-called
Eilenberger
equations~\cite{Eilenberger66,LarkinOvchinnikov,Shelankov,DemlerArnoldBeasley97}
(at the expense of losing detailed information about  excitations)
by introducing the energy-integrated Green functions,
\begin{eqnarray}
\hat{g}(\omega,k,R)\equiv \frac{i}{\pi} \int d\xi_k\,
 \hat{G}(\omega,k,R) ,
 \quad \hat{f}(\omega,k,R)\equiv \frac{1}{\pi} \int d\xi_k\,
 \hat{F}(\omega,k,R), 
\end{eqnarray}
and similarly for $\hat{g}^\dagger(\omega,k,R)$ and
$\hat{f}^\dagger(\omega,k,R)$. 
Here, $\omega$ is the fermionic frequency Fourier conjugate to
the relative time, $k$ is the relative momentum conjugate to the relative coordinate, and $R$ is the center-of-mass
coordinate. (We shall consider stationary processes, so we have
dropped any dependence on the center-of-mass time.)
However,  the resulting equations do not determine $g$'s and $f$'s
uniquely,
and they need to be supplemented by additional normalization conditions~\cite{Eilenberger66,LarkinOvchinnikov,Shelankov,DemlerArnoldBeasley97},
\begin{equation}
\hat{g}^2+ \hat{f} \hat{f}^\dagger = \hat{g}^{\dagger2}+\hat{f}^\dagger \hat{f}=\hat{1},
\end{equation}
as well as the self-consistency equation, 
\begin{equation}
\Delta=|g|\sum\nolimits_\omega f_{12}(\omega).
\label{please-label-me}
\end{equation}

\def\unithat{\check} In the dirty limit (i.e.~$\omega\tau_\text{tr}
\ll G$ and $\Delta\tau_\text{tr} \ll F$), where $\tau_\text{tr}$ is the
transport relaxation time (which we do not distinguish from the
elastic mean-free time), the Eilenberger equations can be simplified
further, because, in this limit, the energy-integrated Green functions
are almost isotropic in  $k$.  This allows one to
retain only the two lowest spherical harmonics ($l=0,1$), and
to regard the $l=1$ term as a small correction
(i.e.~$|\unithat{k}\cdot\vec{F}|\ll |F|$) so that we may write
\begin{equation}
g(\omega,\unithat{k},R)=G(\omega,R)+\unithat{k}\cdot\vec{G}(\omega,R), 
\quad 
f(\omega,\unithat{k},R)=F(\omega,R)+\unithat{k}\cdot\vec{F}(\omega,R),
\end{equation} 
where $\unithat{k}$ is the unit vector along $k$.
In this nearly-isotropic setting, the normalization conditions
simplify to
\begin{align}
G_{11}^2&=1-F_{12}F_{21}^\dagger,&
G_{22}^2&=1-F_{21}F_{12}^\dagger,
\end{align}
and the Eilenberger equations reduce to the celebrated Usadel
equations~\cite{Usadel70} for \(F_{12}(\omega,R)\), \(F_{21}(\omega,R)\),
\(F^\dagger_{12}(\omega,R)\), and \(F^\dagger_{21}(\omega,R)\).

\section{Application to thin wires and films}
Let us consider a wire (or film) not much thicker than the effective
coherence length. In this regime, we may assume that the order
parameter has the form \(\Delta(R)=\tilde{\Delta} e^{i u R_x}\), where
\(R_x\) is the coordinate measured along the direction of the current
(e.g.~for a wire this is along its length) and $u$ is a parameter
encoding the velocity of the superflow \(\hbar u/2m\).  Similarly, we
may assume that the semiclassical anomalous Green functions have a
similar form:
\begin{subequations}
\begin{align}
F_{12}(\omega,R)&=\tilde{F}_{12}(\omega) e^{i u R_x}, 
&
F_{21}(\omega,R)&=\tilde{F}_{21}(\omega) e^{i u R_x}, 
\\
F^\dagger_{12}(\omega,R)
&=\tilde{F}^\dagger_{12}(\omega) e^{-i u R_x}, 
&
F^\dagger_{21}(\omega,R)
&=\tilde{F}^\dagger_{21}(\omega) e^{-i u R_x}.
\end{align}
\end{subequations}
Together with the symmetry amongst $\tilde{F}$'s (i.e.
$\tilde{F}^*_{\alpha\beta}=-\tilde{F}^\dagger_{\alpha\beta}$ and
$\tilde{F}_{\alpha\beta}=-\tilde{F}^*_{\beta\alpha}$)
we can reduce the four Usadel equations for $\tilde{F}_{12}$,
$\tilde{F}_{21}$, $\tilde{F}^\dagger_{12}$, and
$\tilde{F}^\dagger_{21}$ to one single equation:
\begin{align}
&\Bigg[
\omega + i \delta_B +  
\frac{T}{2 \tau_\sf}  
\sum_{\omega'} \Big(D^{-+}_{\omega'} {G}_{22}(\omega -\omega')
\Big)+ 
\Big(\frac{d^z}{\tau_\sf}+\frac{\tilde{D}}{2} \Big)
{G}_{11}(\omega)
+\frac{1}{3 \tau_{\so}} {G}_{22}(\omega) \Bigg]
\frac{\tilde{F}_{12}(\omega)}{\tilde{\Delta}}
\nonumber \\
&\qquad\qquad
- {G}_{11}(\omega)
=
-{G}_{11}(\omega) \frac{T}{2 \tau_\sf} \sum_{\omega'} 
\Big(
D^{-+}_{\omega'} 
\frac{\tilde{F}^*_{12}(\omega-\omega')}{\tilde{\Delta}^*}
\Big)+
\frac{1}{3 \tau_{\so}} 
G_{11}(\omega)\frac{\tilde{F}^*_{12}(\omega)}{\tilde{\Delta}^*},
\label{usadel_main}
\end{align}
in which $\delta_B\equiv \mu_B B+ n_i u_2(0)\overline{S^z}$,
$\tilde{D}\equiv D\dbk{(u-2 e A/c)^2}$ with the London gauge chosen
and $\dbk{\cdots}$ defining a spatial average
over the sample thickness, $D\equiv v_F^2\tau_\text{tr}/3$ is the
diffusion constant. The spin-exchange and spin-orbit scattering times,
$\tau_\sf$ and $\tau_{\so}$, are defined via the Fermi surface
averages
\begin{align}  
\frac{1}{2\tau_\sf} &\equiv N_0 n_i \pi \int\frac{
  d^2\unithat{k}'}{4\pi} |u_2|^2, & \frac{1}{2\tau_{\so}} &\equiv N_0
  n_{\so} \pi \int \frac{d^2\unithat{k}'}{4\pi} |v_{\so}|^2\,
  {p_F^2 |\check{k} \times \check{k}'|^2 }.  
\end{align} 
Here, \(N_{0}\) is the (single-spin) density of electronic states at
  the Fermi surface, \(n_{i}\) is the concentration of magnetic
  impurities, \(n_{\so}\) is the concentration of spin-orbit
  scatterers, and $p_F=m v_F$ is the Fermi momentum.  The normalization
  condition then becomes
\begin{align}
\tilde{G}_{11}(\omega)&=\sgn(\omega)
[{1-\tilde{F}^2_{12}(\omega)}]^{1/2},
& 
\tilde{G}_{22}(\omega)&=\sgn(\omega)
[{1-\tilde{F}^{*2}_{12}(\omega)}]^{1/2}=\tilde{G}_{11}^*(\omega).
\label{root-eqs}  
\end{align}
 Furthermore, the self-consistency condition~(\ref{please-label-me}) becomes 
\begin{eqnarray}
\tilde{\Delta} \ln\big({T_{C0}}/{T}\big) =\pi T{\sum}_{\omega}\Big(\big({\tilde{\Delta}}/{|\omega|}\big)-
\tilde{F}_{12}(\omega)\Big), 
\end{eqnarray}
in which we have exchanged the coupling constant $g$ for $T_{C0}$,
i.e., the critical temperature of the superconductor in the absence of
magnetic impurities and fields.

In the limit of strong
spin-orbit scattering (i.e.~\(\tau_{\so} \ll 1/\omega\) and
\(\tau_\sf\)),  the imaginary part of Eq.~(\ref{usadel_main})
is simplified to 
\begin{subequations}
\begin{eqnarray}
\big[
\delta_B+\frac{T}{2\tau_\sf}
\myIm
{\sum}_{\omega'}D_{\omega'}^{-+}\,G(\omega\!-\!\omega')
\big] \myRe\, C + \frac{2}{3 \tau_{\so}} \myIm(G C)=0, 
\end{eqnarray}
and the real part is rewritten as
\begin{eqnarray}
& \omega \,\myRe\,C
+\frac{T}{2\tau_\sf}
\myRe
\sum_{\omega'} 
\big[
D^{-+}(\omega')\,G(\omega\!-\!\omega')\, C(\omega) +
G^*(\omega)\, D^{-+}(\omega')\,C^*(\omega-\omega')
\big]
\nonumber\\
& -  \big[\delta_B+\frac{T}{2\tau_\sf}
\myIm\sum_{\omega'}D_{\omega'}^{-+}\,G(\omega\!-\!\omega')
\big] \myIm\,C+ 
\left(\frac{d^z}{\tau_\sf}+\frac{\tilde{D}}{2}\right) 
\myRe(G  C) = 
\myRe\, G ,
\end{eqnarray}
\label{Usadel-spin-orbit}
\end{subequations}
where $C\equiv\tilde{F}_{12}/\tilde{\Delta}$,
$G\equiv{G}_{11}$, and the argument $\omega$
is implied for all Green function, except where stated
otherwise. Next, we take the advantage of the simplification that follows
by restricting our attention to the weak-coupling limit, in which
\(\tilde{F}_{12}(\omega)\ll 1\).  Then, eliminating \(G\) in
Eq.~(\ref{Usadel-spin-orbit}) using Eqs.~(\ref{root-eqs}), and
expanding to third order in powers of \(\tilde{F}\), one arrives at an
equation for \(\tilde{F}\) that is readily amenable to numerical treatment.
The quantitative results that we now draw are based on this strategy.
\footnote{We note that, simplifications associated with the  strong
  spin-orbit scattering assumption and the power series expansion in
  $\tilde{F}$ are only necessary to ease the numerical calculations.
Our conclusions are not sensitive to these simplifications in the
  parameter regimes considered in Figs.~\ref{tc_vs_tB} and \ref{jc_vs_tB}. }

\section{Results for the critical temperature\/}
These can be obtained in the standard way, i.e., by (i)~setting $u=0$
and expanding Eqs.~(\ref{Usadel-spin-orbit}) to linear order in
\(\tilde{F}\) (at fixed \(\tilde{\Delta}\)), and (ii)~setting
$\tilde{\Delta} \rightarrow 0$ and applying the self-consistency
condition.  Step~(i) yields
\begin{subequations}
\begin{equation}
\Big[
|\omega|
+\tilde{\Gamma}_\omega
+\frac{D}{2}
\Bigdbk{\Big(\frac{2eA}{c}\Big)^2}
+\frac{3\tau_{\so}}{2}{\delta}_B'(\omega)^2
\Big]
\myRe\, C(\omega)
\approx
1-\frac{T}{\tau_\sf}\sum_{\omega'}\frac{\omega_s\overline{S^z}}{\omega'^2+\omega_s^2}\myRe\, C(\omega-\omega'),
\end{equation}
where 
\begin{equation}
\delta_B'(\omega)
\equiv
\delta_B - 
\frac{T}{\tau_\sf}
\sum_{\omega_c>|\omega'|>|\omega|}\frac{2|\omega'|\overline{S^z}}{{\omega'}^2
+\omega_s^2},
\end{equation}
\end{subequations}
in which  a cutoff $\omega_c$ has been imposed on $\omega'$, and 
\begin{equation}
\tilde{\Gamma}_\omega
\equiv
\frac{d^z}{\tau_\sf}+
\frac{T}{\tau_\sf}\sum_{|\omega'|<|\omega|}
\frac{\omega_s\overline{S^z}}{{\omega'}^2+\omega_s^2}.
\end{equation}
This is essentially the Cooperon equation in the strong spin-orbit scattering limit, first derived by Kharitonov and Feigel'man~\cite{KharitonovFeigelman05}, up to an inconsequential renormalization of $\delta_B$.  

Step~(ii) involves solving the implicit equation
\begin{equation}
\ln\frac{T_{C0}}{T}=
\pi T\sum_{\omega}
\left[
\frac{1}{|\omega|}-
\frac{1}{2}\Big(C(\omega)+C^*(\omega)\Big)
\right], 
\end{equation} 
the solution of which is \(T=T_{C}\).

Figure~\ref{tc_vs_tB} shows the dependence of the critical temperature
of wires or thin films on the (parallel) magnetic field for several
values of magnetic impurity concentration.  Note the qualitative
features first obtained by Kharitonov and
Feigel'man~\cite{KharitonovFeigelman05}: starting at low
concentrations of magnetic impurities, the critical temperature
decreases monotonically with the applied magnetic field.  For larger
concentrations, a marked non-monotonicity develops, and for yet larger
concentrations, a regime is found in which the magnetic field first
induces superconductivity but ultimately destroys it.  The physical
picture behind this is the competition mentioned in the Introduction:
first, by polarizing the magnetic impurities the magnetic field
suppresses their pair-breaking effect.  At yet larger fields, this
enhancing tendency saturates, and is then overwhelmed by the
pair-breaking tendency of the orbital coupling to the magnetic field.

\begin{figure}
\twofigures[width=4.8cm, angle=-90]{figs/tcplot_pretty2.epsi}{figs/jcplot2.epsi}
\caption{\label{tc_vs_tB} Critical temperature vs.~(parallel) magnetic field for
a range of  exchange scattering strengths characterized by
the dimensionless parameter \(\alpha\equiv\hbar/(k_B T_{C0} \tau_{\sf})\).  The strength
for potential scattering is characterized by 
parameter $\hbar/(k_B T_{C0}\tau_\text{tr}) =10000.0$, and that for
the spin-orbit scattering is by $\hbar/(k_B T_{C0} \tau_\so) =
1000.0$; the sample thickness is $ d = 90.0\, \hbar/p_F$, where $p_F$ is
the Fermi momentum; the impurity gyromagnetic ratio is chosen to be
$g_s = 2.0$; and the typical scale of the exchange energy $u_2$
in Eq.~(\ref{Vab}) is
taken to be $E_F/7.5$, where $E_F$ is the Fermi energy. }
\caption{\label{jc_vs_tB} Critical current vs.~(parallel) magnetic
field at several values of temperature, with 
the strength of the exchange scattering set to be
\(\alpha=0.5\)
(corresponding to the solid line in Fig.~\ref{tc_vs_tB}),
and all other parameters being the same
 as used in Fig.~\ref{tc_vs_tB}.}
\end{figure}

\section{Results for the critical current density}
To obtain the critical current density $j_c$, we first determine the
current density (average over the sample thickness) from the solution
of the Usadel equation via
\begin{eqnarray}
j(u)=2eN_0\pi D T{\sum}_{\omega}\myRe\Big(\tilde{F}_{12}^2(\omega)\big[u-\frac{2e}{c}{\dbk{A}}\big]\Big),
\label{current} 
\end{eqnarray}
 and then maximize $j(u)$ with respect to $u$. In the previous
section, we have seen that, over a certain range of magnetic impurity
concentrations, $T_{C}$ displays an upturn with field at small fields,
but eventually decreases.  Not surprisingly, our calculations show
that such non-monotonic behavior is also reflected in the critical current.

Perhaps more interestingly, however, we have also found that for small
concentrations of magnetic impurities, although the critical
temperature displays {\it no\/} non-monotonicity with the field, the
critical current {\it does\/} exhibit non-monotonicity, at least for
lower temperatures.  This phenomenon, which is exemplified in
Fig.~\ref{jc_vs_tB}, sets magnetic impurities apart from other
de-pairing mechanisms.  The reason why the critical current shows
non-monotonicity more readily than the critical temperature does is
that the former can be measured at lower temperatures, at which the
impurities are more strongly polarized by the field.

\section{Conclusion and outlook}
We address the issue of superconductivity, allowing for the
simultaneous effects of magnetic fields and magnetic impurity
scattering, as well as spin-orbit impurity scattering.  In
particular, we investigate the outcome of the two competing roles that
the magnetic field plays: first as a quencher of magnetic impurity
pair-breaking, and second as pair-breaker in its own right.  Thus,
although sufficiently strong magnetic fields inevitably destroy
superconductivity, the interplay between  its two effects 
can, at lower field-strengths, lead to the enhancement of
superconductivity, as first predicted by Kharitonov and
Feigel'man via an analysis of the superconducting transition
temperature.  In the present Letter, we adopt the Eilenberger-Usadel
semiclassical approach, and are thus able to recover the results of
Kharitonov and Feigel'man, which concern the temperature at which the
normal state becomes unstable with respect to the formation of
superconductivity; but we are also able to address the properties of
the superconducting state itself.  In particular, our approach allows
us to compute the critical current and specifically, its dependence on
magnetic field and temperature.

We have found that any non-monotonicity in the field-dependence of the
critical temperature is always accompanied by the non-monotonicity of
the field-dependence of the critical current. However, we have also
found that for a wide range of physically reasonable values of the parameters
the critical current exhibits non-monotonic behavior with
field at lower temperatures, even though there is no such behavior in
the critical temperature.

Especially for small samples, for which thermal fluctuations can smear
the transition to the superconducting state over a rather broad range
of temperatures, the critical current is expected to provide a more
robust signature of the enhancement of superconductivity, as it can be
measured at arbitrarily low temperatures. In addition, the critical
currents can be measured over a range of temperatures, and can
thus provide rather stringent tests of any theoretical models.
Recent experiments measuring the critical temperatures and critical
currents of superconducting MoGe and Nb nanowires show behavior
consistent with the predictions of the present Letter, inasmuch as
they display monotonically varying critical temperatures but
non-monotonically varying critical currents~\cite{RogachevEtAl}.

\acknowledgments 
We acknowledge useful discussions with A.\ J.\
Leggett and M.\ Yu.\ Kharitonov.  This work was supported via NSF
EIA01-21568 and CAREER award DMR01-34770, via DOE DEFG02-91ER45439,
and via the A.\ P.\ Sloan Foundation.


\end{document}